\documentclass[%
superscriptaddress,
a4paper,
amsmath,amssymb,
aps,
prd,
]{revtex4-2}

\usepackage{dcolumn}
\usepackage{bm}
\usepackage{graphicx}
\usepackage[colorlinks, linkcolor=blue, anchorcolor=blue, citecolor=blue]{hyperref}
\usepackage{subfigure}
\usepackage{multirow}
\usepackage{acronym}
\usepackage{tabularx}
\usepackage{booktabs}

\usepackage{color}

\begin{document}

\title{Parameter inference for coalescing massive black hole binaries using deep learning}

\author{Wen-Hong Ruan}
\email{ruanwenhong@ucas.ac.cn}
\affiliation{School of Fundamental Physics and Mathematical Sciences, Hangzhou Institute for Advanced Study, University of Chinese Academy of Sciences, Hangzhou 310024, China}
\affiliation{School of Physical Sciences, University of Chinese Academy of Sciences, No.19A Yuquan Road, Beijing 100049, China}

\author{He Wang}
\email{hewang@ucas.ac.cn}
\affiliation{International Centre for Theoretical Physics Asia-Pacific, University of Chinese Academy of Sciences, Beijing 100049, China}
\affiliation{Taiji Laboratory for Gravitational Wave Universe, University of Chinese Academy of Sciences, Beijing 100049, China}

\author{Chang Liu}
\email{liuchang@ucas.ac.cn}
\affiliation{School of Fundamental Physics and Mathematical Sciences, Hangzhou Institute for Advanced Study, University of Chinese Academy of Sciences, Hangzhou 310024, China}
\affiliation{School of Physical Sciences, University of Chinese Academy of Sciences, No.19A Yuquan Road, Beijing 100049, China}

\author{Zong-Kuan Guo}
\email{guozk@itp.ac.cn}
\affiliation{CAS Key Laboratory of Theoretical Physics, Institute of Theoretical Physics, Chinese Academy of Sciences, Beijing 100190, China}
\affiliation{School of Physical Sciences, University of Chinese Academy of Sciences, No.19A Yuquan Road, Beijing 100049, China}
\affiliation{School of Fundamental Physics and Mathematical Sciences, Hangzhou Institute for Advanced Study, University of Chinese Academy of Sciences, Hangzhou 310024, China}

\begin{abstract}
In the 2030s, a new era of gravitational-wave (GW) observations will dawn as multiple space-based GW detectors, such as the Laser Interferometer Space Antenna, Taiji and TianQin, open the millihertz window for GW astronomy.
These detectors are poised to detect a multitude of GW signals emitted by different sources.
It is a challenging task for GW data analysis to recover the parameters of these sources at a low computational cost.
Generally, the matched filtering approach entails exploring an extensive parameter space for all resolvable sources, incurring a substantial cost owing to the generation of GW waveform templates.
To alleviate the challenge, we make an attempt to perform parameter inference for coalescing massive black hole binaries (MBHBs) using deep learning.
The model trained in this work has the capability to produce 50,000 posterior samples for redshifted total mass, mass ratio, coalescence time and luminosity distance of a MBHB in about twenty seconds.
Our model can serve as a potent data pre-processing tool, reducing the volume of parameter space by more than four orders of magnitude for MBHB signals with a signal-to-noise ratio larger than 100.
Moreover, the model exhibits robustness when handling input data that contains multiple MBHB signals.
\end{abstract}

\maketitle

\section{Introduction}
\label{sec:Intro}
Several space-based gravitational-wave (GW) detectors are expected to launch in the 2030s, including the Laser Interferometer Space Antenna (LISA)~\cite{amaro2017laser}, Taiji~\cite{hu2017taiji} and TianQin~\cite{luo2016tianqin}, which will conduct the all-sky survey of GWs in millihertz frequency band.
One of the key focal points for these detectors is the GW signals emitted by coalescing massive black hole binaries (MBHBs) with total masses ranging from $10^5 M_\odot$ and $10^8 M_\odot$.
Based on the estimations from population models, it is expected to detect more than one MBHB coalescences per year~\cite{klein2016science}.
It is a key task for space-based GW data analysis to recover the parameters of these systems.
By harnessing this information, we can trace the origin, growth and merger history of MBHBs.

Typically, the matched filtering (MF) method~\cite{owen1999matched} stands as the primary choice for analysing a weak signal buried in noise.
It has been widely used to infer the parameters of stellar-mass binary black holes (BBHs) for ground-based GW detection~\cite{allen2012findchirp,abbott2016ligo,abbott2016gw151226}.
However, the method is computationally expensive due to extensive generation of waveform templates during the stochastic exploration of parameter space.
As the number of GW events detected by LIGO/Virgo Collaboration continues to surge, the substantial cost of the MF method becomes increasingly undeniable.
The challenge further intensifies when applying the MF method to space-based GW detection.
Because the GW waveform templates are more complicated due to the motion of detectors and the application of time-delay interferometry (TDI) technique~\cite{tinto2014time}.
Moreover, GW signals emitted by some sources can be observed for several days, weeks or even months during the lifetime of a detector.
It is the case for MBHBs considered in this work.
It is foreseeable that the strain data of the detector will contain a mixture of multiple GW signals from different sources.
Generally, a global-fit analysis has been proposed to recover the source parameters of all resolvable signals~\cite{cornish2005lisa,littenberg2020global,littenberg2023prototype}.
The method explores a large parameter space spanned by the parameters of all sources, resulting in a considerable computational cost.
As the launch time of the space-based GW detectors approaches, it is urgent and necessary to develop novel techniques that can effectively mitigate the computational cost of parameter inference.

Currently, the application of deep learning in parameter inference has garnered significant attention within the GW community.
Many researchers have applied deep learning models to produce the posterior for source parameters of stellar-mass BBHs~\cite{george2018deep,green2020gravitational,krastev2021detection,green2021complete,dax2021real,shen2021statistically,schmidt2021machine,gabbard2022bayesian,langendorff2023normalizing}.
Some of the models can achieve comparable performance with the MF approach on the GW events detected by LIGO/Virgo.
Moreover, some authors have considered the parameter inference for space-based GW detection through deep learning models in~\cite{chua2020learning}.
The authors implement a successful example of producing two-dimensional posterior for MBHBs with components masses $m_{1,2} \in [1.25, 10] \times 10^5 M_{\odot}$ by a deep learning model.
Their model is trained on the family of 2.5 PN TaylorF2 waveforms which are characterized by five parameters: masses and spins of the two black holes, as well as the signal-to-noise ratio (SNR) of the waveform.
Moreover, the detector responses of TDI channels are not considered.

In this paper, we present an implementation of parameter inference for nonprecessing spinning MBHBs using deep learning model based on the normalizing flow (NF)~\cite{rezende2015variational}.
Actually, the NF architecture has demonstrated remarkable capability in parameter inference for ground-based GW sources~\cite{green2021complete,dax2021real,shen2021statistically,langendorff2023normalizing}.
In light of this success, we make an attempt to extend its application to coalescing MBHBs detected by future space-based GW detectors.
Taking the simulated LISA data as input, our model has the ability to produce reliable posterior for redshifted total mass, mass ratio, coalescence time and luminosity distance of MBHBs with instrumental noise.
The model takes only about twenty seconds to draw 50,000 posterior samples for the four parameters, which is much faster than the MF approach.
Although our model provides less precise ranges of the parameters compared to the MF approach, we introduce an attempt on the parameter inference for coalescing massive black hole binaries. %
It serves as valuable data pre-processing.
Specifically, the model rapidly establishes a narrowed prior within tens of seconds, which is negligible when contrasted with the time-consuming random sampling in the parameter space.
And the computational cost of the MF can be reduced by adopting the narrowed prior.
Furthermore, our model shows robustness against the presence of multiple MBHB signals in strain data.
This noteworthy characteristic indicates it as a potential candidate for integration with the global-fit analysis, enabling a lower computational cost to recover the source parameters of all resolvable MBHB signals.
In this work, we use simulated LISA data to train our model. In principle, it can be easily extended to other space-based GW detectors.

The paper is organized as follows. In Sec.~\ref{sec:Method}, we introduce the framework of our model. In Sec.~\ref{sec:Datasets}, we illustrate the generation of simulated LISA data used to train and test the model. Next, Sec.~\ref{sec:Results} shows the test results of our model. Finally, we give summary and discussions in Sec.~\ref{sec:Discussion}.

\section{model}
\label{sec:Method}
\begin{figure}[htb]
    \centering
    \includegraphics[width=1\linewidth]{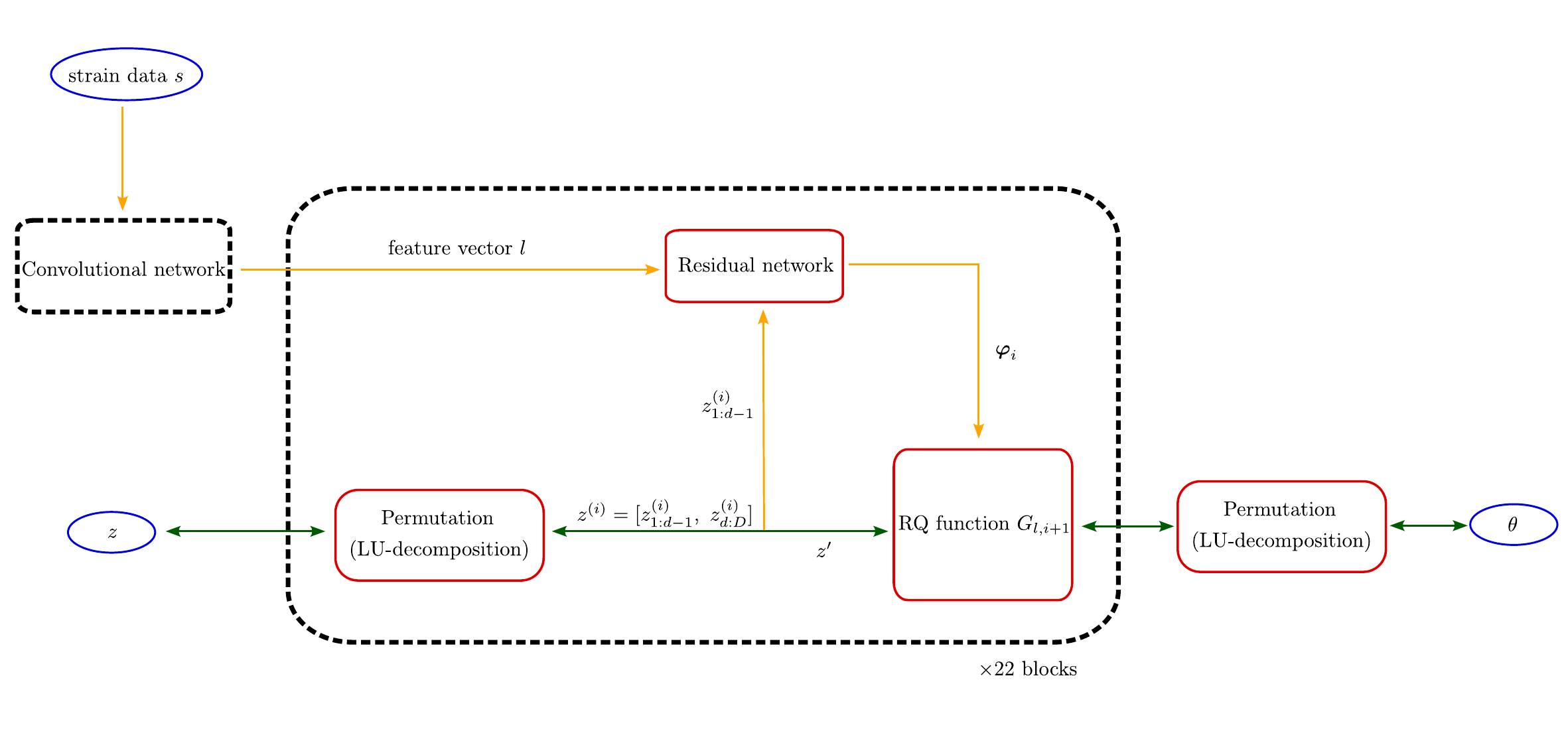}
    \caption{Framework of the neural network used in this work. The small black dashed box represents the CNN used to extract key features from the strain data $s$. The large black dashed box depicts the block of the NF and the number of the blocks is set to 22. The initial variable $z$ is transformed to $\theta$ after passing through the whole neural network.}
    \label{fig:flow}
\end{figure}

Our task is to obtain the posterior $p(\Theta|s)$ of the source parameters $\Theta$ from the strain data $s$.
To achieve the target, we construct a generative model conditioned on the strain data to draw samples of a random variable $\theta$.
In other words, the model is a sample generator of conditional distribution $q(\theta|s)$.
By tuning the learnable parameters of the model during training, the conditional distribution transforms into an estimation of the posterior for source parameters.
Practically, we combine a convolutional neural network (CNN)~\cite{lecun1998gradient} and an NF to construct the model.
The framework of our model is shown in Fig.~\ref{fig:flow}, which can be divided into two parts.

The first part of the model is designed to extract key features from the strain data $s$ which is indicated in Fig.~\ref{fig:flow} with the smaller black dashed box.
This part is required to encode the features into a low-dimensional output, as a higher-dimensional output would lead to an increase in the parameters of the subsequent network, consequently raising the cost of the training process.
We have explored several neural network structures commonly used in GW community to implement this part,  and after careful consideration, we settled on the CNN due to its final performance and training difficulty.
Specifically, the CNN is composed of 3 convolutional layers and 3 fully connected layers.
It encodes the key information of the strain data in a feature vector $l$ to input the rest part of the model.
The dimension of $l$ is set to 256 in this work.

The second part of the model is implemented by an NF, which describes the transformation of an initial probability distribution into another probability distribution.
The transformation is defined as an invertible and smoothing mapping $f_l: R^D \rightarrow R^D$, where D is the dimension of sample space.
In this paper, we set $D=4$ for redshifted total mass, mass ratio, coalescence time and luminosity distance of MBHB.
Note that, the NF is conditioned on the feature vector $l$.
Applying the transformation on a random variable $z$ with distribution $\pi(z)$, the resulting random variable $\theta = f_l(z)$ obeys the distribution~\cite{rezende2015variational}
\begin{equation}
    \label{eq:q_f}
    q(\theta|s) = \pi(z)\left| \det \left(\frac{\partial f_l}{\partial z}\right)\right|^{-1}.
\end{equation}
The transform comprises a series of artificially defined invertible and smoothing mappings, which can be written as
\begin{equation}
    \label{eq:f_l}
    f_l(z) = f_{l,N} \circ f_{l,N-1} \circ \cdots \circ f_{l,1}(z).
\end{equation}
Each mapping $f_{l, i} \ (i=1,\cdots,N)$ represents a block of the NF which correspond to the bigger black dashed box in Fig.~\ref{fig:flow}.
There are many types of NF due to different designs of the mapping $f_{l, i}$.
In this work, we adopt a neural spline flow~\cite{durkan2019neural} that is also used in some researches of ground-based GW data analysis~\cite{green2021complete,dax2021real,shen2021statistically}.
The construction of the neural spline flow block is depicted within the bigger black dashed box in Fig.~\ref{fig:flow}.
For the input $z^{(i)} = f_{l,i} \circ \cdots \circ f_{l,1}(z)$ of the $(i+1)$th block, it undergoes an initial step of random permutation through an intermediate layer which is implemented by LU-decomposition approach~\cite{oliva2018transformation}.
The layer ensures the components of $z^{(i)}$ can interact with others.
Then the output of the permutation layer is split into two parts, which can be written as
\begin{eqnarray}
    z^{(i)} = [z^{(i)}_{1:d-1}, z^{(i)}_{d:D}].
\end{eqnarray}
Next, $z^{(i)}$ is inputted to a monotonic rational-quadratic (RQ) function $G_{l, i}$ which transform the two parts of $z^{(i)}$ separately by
\begin{equation}
    G_{l, i}(z^{(i)})= \begin{cases}z^{(i)}_j & \text { if } j < d, \\ g_{\varphi_i}\left(z^{(i)}_j\right) & \text { if } j \geq d,\end{cases}
\end{equation}
where $g_{\varphi_i}$ is an artificially defined function parameterized by $\varphi_i$ and the parameters are determined by $l$ and $z^{(i)}_{1:d-1}$.
Specifically, the function $g_{\varphi_i}$ maps an interval $[-L, L]$ to $[-L, L]$ and divides the interval into $K$ bins (see more details in Fig.1 of~\cite{durkan2019neural}).
Each bin is continuously separated by $K + 1$ coordinates $\{(x^{(k)}, y^{(k)}) \}_{k=0}^K$ with $(x^{(0)}, y^{(0)}) = (-L, -L)$ and $(x^{(K)}, y^{(K)}) = (L, L)$.
For $x \in [-L, L]$, the function in the $k$th bin is evaluated through a monotonically-increasing function which is given by~\cite{durkan2019neural}
\begin{eqnarray}
    g_{\varphi_i}^{(k)}(\xi) &=& y^{(k)}+\frac{\left(y^{(k+1)}-y^{(k)}\right)\left[\tau^{(k)} \xi^2+\delta^{(k)} \xi(1-\xi)\right]}{\tau^{(k)}+\left[\delta^{(k+1)}+\delta^{(k)}-2 \tau^{(k)}\right] \xi(1-\xi)}, \\
    \xi (x) &=& \frac{x - x^{(k)}}{x^{(k+1)} - x^{(k)}}, \\
    \tau^{(k)} &=& \frac{y^{(k+1)} - y^{(k)}}{x^{(k+1)} - x^{(k)}},
\end{eqnarray}
where $\delta^{(k)}$ denotes the derivative of the function at coordinate $(x^{(k)}, y^{(k)})$ and the derivatives of boundaries $\{\delta^{(0)}, \delta^{(K)}\}$ are set to 1.
Therefore, the function $g_{\varphi_i}$ is parameterized by $3K - 1$ parameters which can be written as
\begin{equation}
    \varphi_i=\left[\varphi_i^w, \varphi_i^h, \varphi_i^d\right].
\end{equation}
Here, $2K$ of the parameters $\varphi_i^w$ and $\varphi_i^h$ determine the widths and heights of the $K$ bins, respectively.
The rest $K-1$ of the parameters $\varphi_i^d$ determine the boundary derivatives $\{ \delta^{(k)}\}_{k=1}^{K-1}$.
For the neural spline flow used in this work, the parameters $\varphi_i$ are given by a residual network~\cite{he2016deep} which take the feature vector $l$ and the $z^{(i)}_{1:d-1}$ part as input.
The residual network contains 14 residual blocks and each block is combined with two fully-connected hidden layers of 512 units.
Moreover, we set the number of the block of the NF as $N=22$ and the number of bins as $K=8$.
As shown in Fig.~\ref{fig:flow}, the initial variable $z$ is transformed to the variable $\theta$ after passing through all blocks of the NF and a permutation layer.
Generally, it is inclined to choose a simple distribution $\pi(z)$ that is convenient for drawing samples of the variable $z$.
We take $\pi(z)$ as the standard multivariate normal distribution in this work.

To make the distribution $q(\theta|s)$ described by the model close to the GW posterior $p(\Theta|s)$, we train the model by the cross entropy between the two distributions, which is given by~\cite{green2021complete}
\begin{equation}
    \label{eq:H_pq}
    H(p,q) = - \int ds \, p(s) \int d\theta \, p(\theta|s)\log q(\theta|s).
\end{equation}
Practically, we train the model by minimizing the cross entropy which is a metric of the difference between two distributions.
However, it is difficult to obtain the posterior $p(\theta|s)$ in the integral.
The posterior can be converted to likelihood by Bayes' theorem, then the cross-entropy can be written as
\begin{equation}
    H(p,q) = - \int d\theta \, p(\theta) \int ds \, p(s|\theta)\log q(\theta|s).
\end{equation}
In the training stage, $H(p,q)$ should be calculated on every minibatch and the integral can be estimated by a Monte Carlo approximation~\cite{green2021complete,shen2021statistically}
\begin{equation}
    H(p,q) \approx -\frac{1}{B} \sum_{n=1}^{B} \ln q(\theta_n|s_n),
\end{equation}
where $B$ denotes the batch size and $s_n$ represents the $n$th simulated strain data on the minibatch.
Note that, $\theta_n$ is drawn from the prior of source parameters and $s_n$ is generated by the composition of GW waveform and noise, which will be explained in detail in Sec.~\ref{sec:Datasets}.
Furthermore, we use the Adam optimizer~\cite{kingma2014adam} to minimize the cross entropy stochastically on minibatches.
In this work, we implement the model based on PyTorch~\cite{NEURIPS2019_9015}, nflows~\cite{nflows} and codes shared in~\cite{lfigw}.

\section{Datasets}
\label{sec:Datasets}
The strain data $s$ considered in this work is composed of GW signal and detector noise, which can be written as
\begin{equation}
    \label{eq:data}
    s = h(\Theta) + n,
\end{equation}
where $h(\Theta)$ is a GW signal from MBHB with parameters $\Theta$ and $n$ is the detector noise.
We simulate GW signals used in the training and test stage by the IMRPhenomD waveforms~\cite{husa2016frequency,khan2016frequency}, which models nonprecessing spinning inspiral-merger-ringdown waveforms.
The GW waveforms are generated with random sampling over 11-dimensional set of source parameters: redshifted total mass $M$, mass ratio $q$, coalescence time $t_c$, luminosity distance $d_L$, dimensionless spins $(s_{1z}, s_{2z})$, inclination angel $\iota$, ecliptic latitude $\beta$, ecliptic longitude $\lambda$, reference phase $\phi_c$ and polarization angle $\psi$.
The priors of the source parameters are listed in Tab.~\ref{tab:prior}.
The range of coalescence time covers nearly a whole year and the range of luminosity distance is converted from the range of redshifts $z \in [0.5, 5]$ assuming a flat $\mathrm{\Lambda CDM}$ cosmology with $\Omega_m=0.31,\Omega_\Lambda=0.69$ and $H_0=67.74$~\cite{ade2016planck}.
Moreover, to simulate the GW signal in real data, the IMRPhenomD waveforms should be modulated by the response function.
For LISA, one of the strongest components of instrumental noise is the laser phase noise.
It can be suppressed by the TDI technique which combines measurements from different arms of LISA into a composite observable~\cite{tinto2014time}.
We choose the uncorrelated TDI observables A and E~\cite{prince2002lisa} as two channels of the input data passing through our model.
Moreover, for the random noise $n$ in strain data, we generate Gaussian instrumental noises of A and E channels by the power spectral density stated in the LISA Science Requirement Document~\cite{lisa_sci_rs}.
In this work, the simulated GW signals are generated by codes of LISA Data Challenge Group (LDC)~\cite{ldcmanual001} and the instrumental noises are generated by PyCBC package~\cite{alex_nitz_2021_5347736}.

For each TDI channel, we set the length of the input data as 98,304 and the sampling rate as $1/5$ Hz.
During each epoch of the training process, we generate 60,000 GW signals by randomly drawing source parameters from the prior and combining them with random noises.
This approach effectively prevents overfitting and enhances the robustness of the model.
As is standard practice in training deep learning models, the training dataset is split into two sets for training and validation.
Specifically, 90\% of the data is allocated for training, while the remaining data is utilized for validation.

\begin{table*}[h]
    \centering
    \setlength{\tabcolsep}{7mm}{
    \begin{tabular}{cc}
        \toprule \hline\hline
        Parameter & Prior \\ \hline
        \midrule
        $M$   & $\text{LogUniform}[10^6 M_{\odot}, 10^7 M_{\odot}]$   \\
        $q$     & $\text{Uniform}[1, 5]$   \\
        $t_c$   & $\text{Uniform}[3 \text{d}, 365 \text{d}]$   \\
        $d_L$   & $\text{Uniform}[2910 \text{Mpc}, 47312 \text{Mpc}]^3$   \\
        $(s_{1z}, s_{2z})$ & $\text{Uniform}[-1, 1]$   \\
        $\cos \iota$ & $\text{Uniform}[-1, 1]$   \\
        $\sin \beta$ & $\text{Uniform}[-1, 1]$   \\
        $\lambda$ & $\text{Uniform}[0, 2\pi]$   \\
        $\phi_c$ & $\text{Uniform}[0, 2\pi]$   \\
        $\psi$ & $\text{Uniform}[0, \pi]$      \\ \hline\hline
        \bottomrule
    \end{tabular}}
    \caption{Priors of the source parameters used in this work.}
    \label{tab:prior}
\end{table*}

\section{Results}
\label{sec:Results}
In this work, we trained many models on an NVIDIA Tesla A40 GPU, experimenting with different learning rates, batch sizes and hyperparameters.
We evaluated the performance of these models based on their final validation loss.
The test results reported in this paper correspond to the model that achieves the lowest validation loss.
The model was trained for about 5 days, employing a batch size of 512 and a total epochs of 3,200.
Moreover, the learning rate is set to 0.0001 at the beginning, which gradually decreases by cosine annealing~\cite{loshchilov2016sgdr} in the training process.
The model takes only about twenty seconds to produce 50,000 posterior samples of the redshifted total mass, mass ratio, coalescence time and luminosity distance for an MBHB signal.

\begin{figure}[htb]
    \centering
    \includegraphics[width=0.5\linewidth]{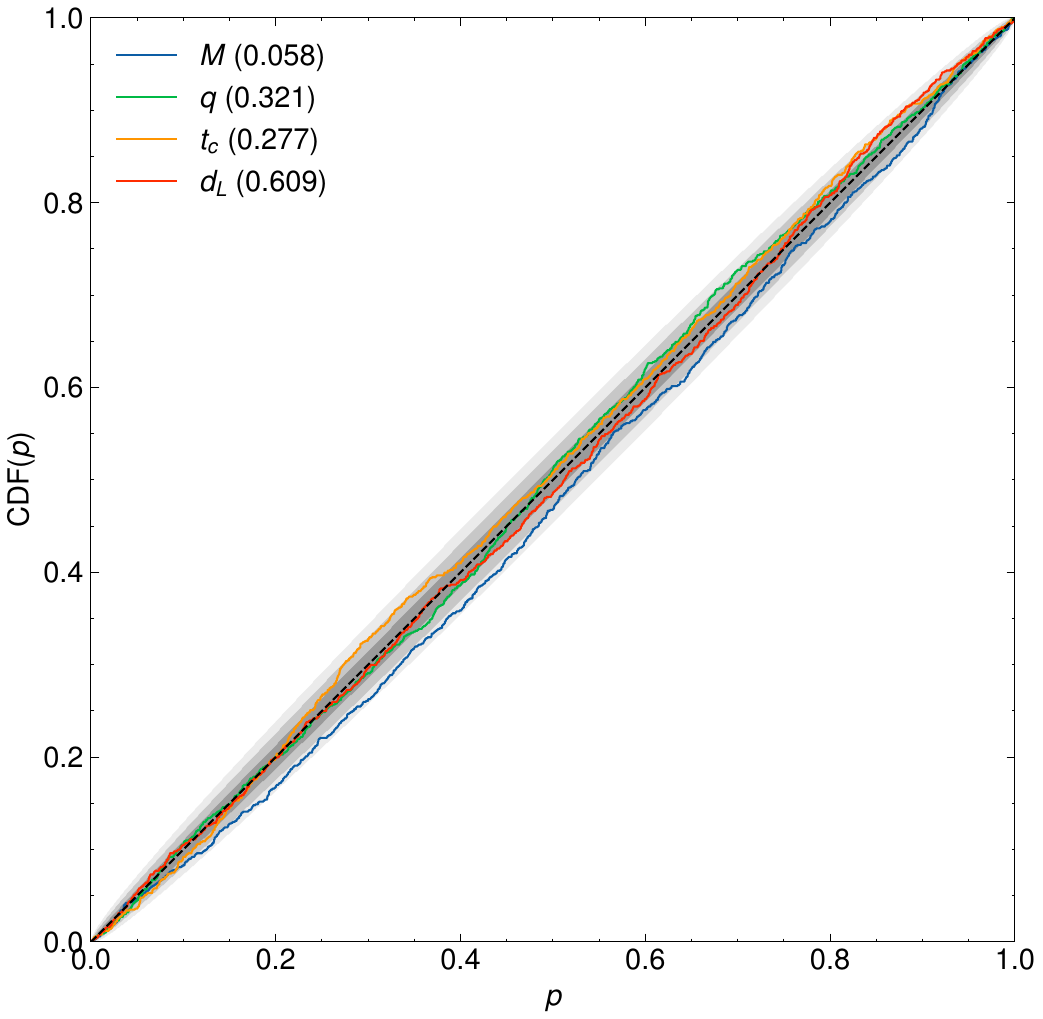}
    \caption{P-P plot for redshifted total mass $M$ (blue), mass ratio $q$ (green), coalescence time $t_c$ (orange) and luminosity distance $d_L$ (red) of MBHB. We generate a test dataset composed of 1,000 simulated strain data to conduct the KS test for our model and the p-values are listed in the upper left corner. The model produces 20,000 samples for each test data. The coloured lines denote the empirical CDF provided by the model and the black dashed line denotes the true CDF. The grey regions represent the $1\sigma$, $2\sigma$ and $3\sigma$ confidence bounds.}
    \label{fig:ks_test}
\end{figure}

Generally, it is convenient to demonstrate the reliability of a generative model through Kolmogorov-Smirnov (KS) test~\cite{massey1951kolmogorov}.
We conduct the KS test on our model using a dataset containing 1,000 simulated strain data.
The GW signals injected in the strain data are generated with source parameters randomly drawn from the prior in Tab.~\ref{tab:prior}.
For each test signal, our model produces 20,000 samples to estimate the posterior for the four source parameters.
The results of the test are summarized in the P-P plot shown in Fig.~\ref{fig:ks_test}.
The p-values of the four parameters are given in the upper left corner of the figure.
The coloured lines in the figure represent the empirical cumulative distribution function (CDF) of the number of times of the true value for each parameter falling within a credible interval $p$, as a function of $p$.
As shown in the figure, the empirical CDF lines provided by our model lie close to the true CDF line (the diagonal black dashed line), which confirms that our model is a reliable estimator of the GW posterior.

\begin{figure}[htb]
    \centering
    \includegraphics[width=0.5\linewidth]{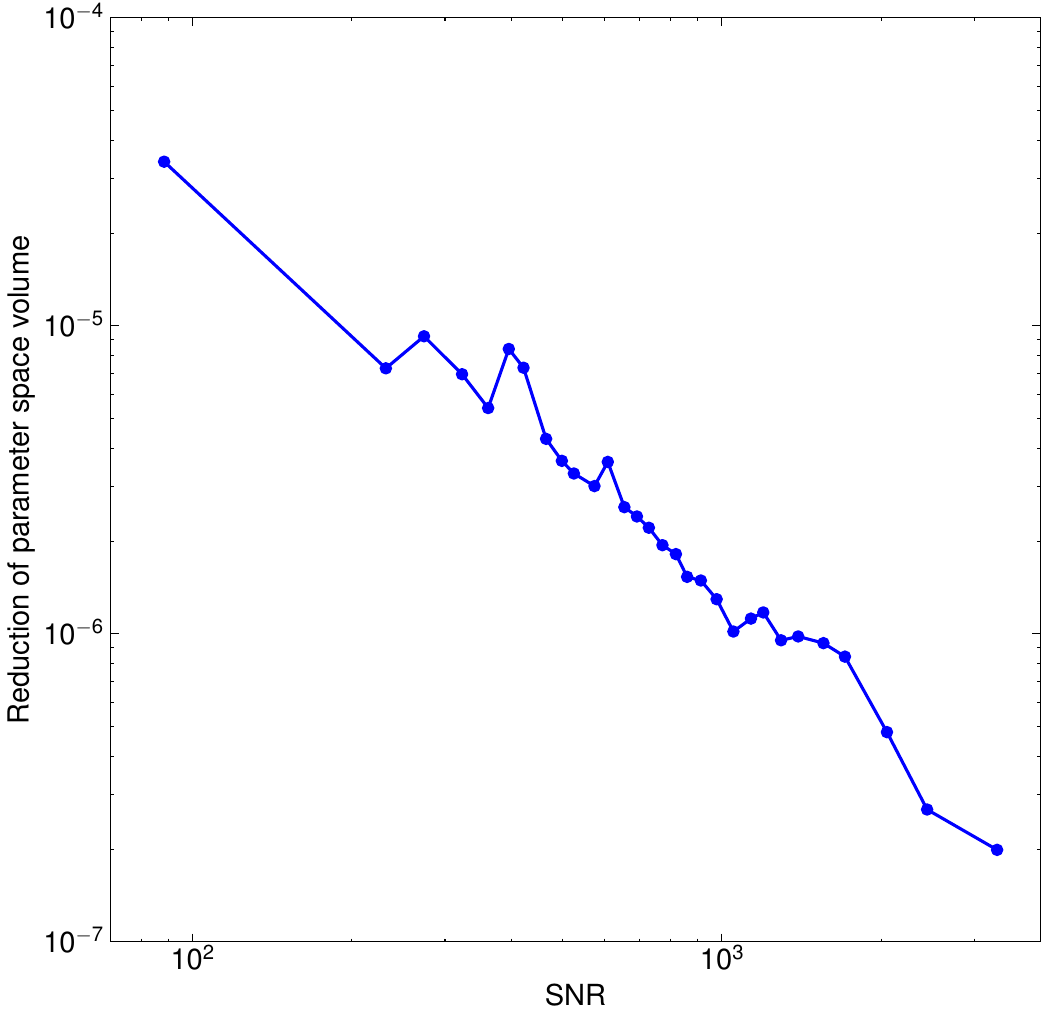}
    \caption{Reduction of parameter space volume as a function of SNR. The value of reduction is calculated based on 90\% credible region obtained from the posterior samples produced by our model in the KS test. The blue line denotes the average value calculated for each bin of SNR.}
    \label{fig:reduction}
\end{figure}

Compared with the uncertainties of source parameters given by the MF approach, our model provides rougher estimations of the source parameters.
Nonetheless, the model can produce an estimated posterior in tens of seconds, which is valuable as a data pre-processing.
Based on the output of our model, we can determine a substantially narrowed prior in comparison to the initial prior used in the MF approach.
The additional time cost incurred by our model is negligible when compared with the cost on the stochastic exploration of parameter space.
Fig.~\ref{fig:reduction} shows the reduction of parameter space volume with the help of our model, as a function of SNR.
Specifically, we estimate the value based on 90\% credible region obtained from the posterior samples produced by our model in the KS test.
The initial priors for $M$, $q$ and $d_L$ used in the MF approach are assumed to be the same as the priors in Tab.~\ref{tab:prior}.
And, the initial prior for $t_c$ is assumed to be the input length of our model.
We average the estimated reductions of parameter space volume for each bin of SNRs.
As shown in Fig.~\ref{fig:reduction}, our model provides a greater reduction for the MBHB with larger SNR.
The model is capable of reducing the parameter space volume by more than four orders of magnitude when $\text{SNR} > 100$.

\begin{figure}[htb]
    \centering
    \includegraphics[width=0.7\linewidth]{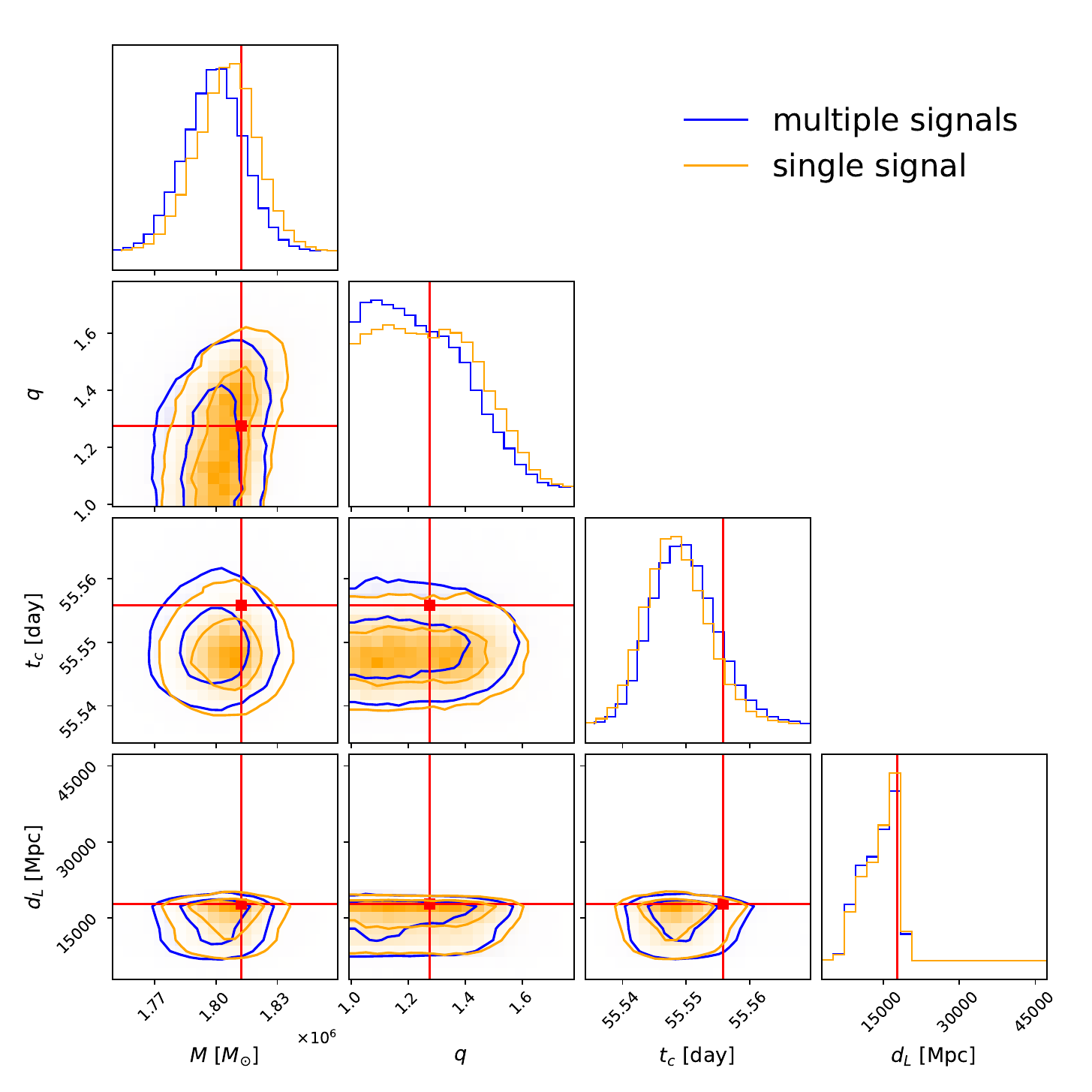}
    \caption{Posterior distributions of the redshifted total mass $M$, mass ratio $q$, coalescence time $t_c$ and luminosity distance $d_L$ produced by our model. The orange line denotes the result for a single MBHB signal and the blue line denotes the result for the same signal overlapped with the inspiral phases of other 14 MBHB signals. The model produces 50,000 posterior samples for each case.}
    \label{fig:overlap}
\end{figure}

Considering that a GW signal emitted by MBHB can evolve in the sensitive frequency band of LISA for days, months or even years, it is foreseeable that multiple MBHB signals will be superimposed in real data.
Although our model is trained on the strain data containing a single MBHB signal, it also exhibits robustness when dealing with the data that contains multiple signals.
The LDC group simulated one year of LISA data comprising a mixture of 15 MBHB signals, with these systems undergoing mergers at different times throughout the year~\cite{ldcweb}.
We compose the mixture signal and instrumental noise to test the performance of our model on multiple signals.
Note that, eight of the MBHBs possess parameters falling within the priors listed in Tab.~\ref{tab:prior}, whereas the remaining MBHBs have luminosity distances beyond the specified range.
Thus, the test for our model is focused on the eight MBHBs which are labelled by LDC with numbers $\{1, 2, 3, 8, 9, 10, 13, 14, 15\}$.
As a comparison, we additionally generate eight GW signals individually using the parameters provided by LDC for these systems.
We then combine each signal with its respective noise separately to test the performance of our model.
Due to the input length limitations of the model, the merger phase of MBHB 2 always exists in input data with the merger phase of another MBHB.
The model cannot produce reliable samples for the input data containing multiple merger phases.
For the other seven MBHBs, our model produces similar posteriors both in single signal and multiple signal cases.
Fig.~\ref{fig:overlap} shows the estimated posteriors of the MBHB 14 produced by the model in these two scenarios.
The MBHB 14 is the first one to evolve to the merger phase, which means the GW signal from it is overlapped with the inspiral phases of all other GW signals.
As shown in Fig.~\ref{fig:overlap}, the marginalized one- and two-dimensional posterior distributions of the multiple signal case (blue) closely align with those of the single signal case (orange).
Our model has the capability to adapt effectively when multiple MBHB signals are observed during the lifetime of LISA.
It is expected that the model can contribute to the global fit analysis by determining a reduced parameter space.

\section{Summary and discussions}
\label{sec:Discussion}

In this paper, we present the implementation of parameter inference using deep learning for coalescing MBHBs.
While training a deep learning model may require some time, taking several days in this study, it can produce a large number of posterior samples of source parameters in a very short duration.
Our model has the ability to generate 50,000 samples of the four source parameters in about twenty seconds.
Due to the complexity of the strain data raised from the motion of space-based detectors and TDI technique, the current model cannot achieve comparable precision of parameter inference when compared with the MF approach.
However, it remains feasible to treat the model as a data pre-processing tool that effectively reduces the parameter space volume by over four orders of magnitude for MBHB signals with $\text{SNR} > 100$.
This reduction consequently reduces the computational cost of the follow-up exploration of parameter space accordingly.

In reality, the real data of space-based GW detection may contain many unexpected features.
Therefore, the current simulated data cannot be completely consistent with future real-world scenarios.
However, the generalization ability of deep learning enables it to work beyond the scope of training data, making it adaptable to handle unseen variations and unexpected features that may arise in the future.
Although our model is trained on data containing a single MBHB signal, it exhibits robustness when analysing data containing multiple MBHB signals.
In our test, the model yields similar posteriors for both scenarios, showcasing the potential of deep learning to be integrated with a global fit analysis for analysing real LISA data.
This combined approach holds promising prospects for handling numerous resolvable GW signals at a low computational cost.

The generalization capability of the deep learning model can be further extended.
In this work, we train the model on Gaussian instrumental noise.
Nevertheless, during real detection, other noise components will be present, such as non-Gaussian and non-stationary foreground noise composed of tens of millions GW signals from Galactic binaries, as well as data gaps and glitches~\cite{cornish2022low}.
Currently, our model has not been extended to analyse input data containing these components.
It can be handled by artificially adding these components to the training data and increasing the complexity of the neural network.
Due to the superior ability of deep learning in dealing with unknown data features, it holds great potential for further development in space-based GW data analysis.

\begin{acknowledgments}

WHR is supported by the National Natural Science Foundation of China under Grant No. 12247140.
CL is supported by the National Natural Science Foundation of China Grant No. 12147132.
ZKG is supported in part by the National Natural
Science Foundation of China Grants No. 12075297 and No. 12235019.
We thank the LDC group to provide the software and datasets.
We also thank the authors of~\cite{green2021complete} for open source codes.

\end{acknowledgments}

\bibliographystyle{apsrev4-2}
\bibliography{pe_flow}

\providecommand{\noopsort}[1]{}\providecommand{\singleletter}[1]{#1}%
\begin{thebibliography}{42}%
\makeatletter
\providecommand \@ifxundefined [1]{%
 \@ifx{#1\undefined}
}%
\providecommand \@ifnum [1]{%
 \ifnum #1\expandafter \@firstoftwo
 \else \expandafter \@secondoftwo
 \fi
}%
\providecommand \@ifx [1]{%
 \ifx #1\expandafter \@firstoftwo
 \else \expandafter \@secondoftwo
 \fi
}%
\providecommand \natexlab [1]{#1}%
\providecommand \enquote  [1]{``#1''}%
\providecommand \bibnamefont  [1]{#1}%
\providecommand \bibfnamefont [1]{#1}%
\providecommand \citenamefont [1]{#1}%
\providecommand \href@noop [0]{\@secondoftwo}%
\providecommand \href [0]{\begingroup \@sanitize@url \@href}%
\providecommand \@href[1]{\@@startlink{#1}\@@href}%
\providecommand \@@href[1]{\endgroup#1\@@endlink}%
\providecommand \@sanitize@url [0]{\catcode `\\12\catcode `\$12\catcode `\&12\catcode `\#12\catcode `\^12\catcode `\_12\catcode `\%12\relax}%
\providecommand \@@startlink[1]{}%
\providecommand \@@endlink[0]{}%
\providecommand \url  [0]{\begingroup\@sanitize@url \@url }%
\providecommand \@url [1]{\endgroup\@href {#1}{\urlprefix }}%
\providecommand \urlprefix  [0]{URL }%
\providecommand \Eprint [0]{\href }%
\providecommand \doibase [0]{https://doi.org/}%
\providecommand \selectlanguage [0]{\@gobble}%
\providecommand \bibinfo  [0]{\@secondoftwo}%
\providecommand \bibfield  [0]{\@secondoftwo}%
\providecommand \translation [1]{[#1]}%
\providecommand \BibitemOpen [0]{}%
\providecommand \bibitemStop [0]{}%
\providecommand \bibitemNoStop [0]{.\EOS\space}%
\providecommand \EOS [0]{\spacefactor3000\relax}%
\providecommand \BibitemShut  [1]{\csname bibitem#1\endcsname}%
\let\auto@bib@innerbib\@empty
\bibitem [{\citenamefont {Amaro-Seoane}\ \emph {et~al.}(2017)\citenamefont {Amaro-Seoane}, \citenamefont {Audley}, \citenamefont {Babak}, \citenamefont {Baker}, \citenamefont {Barausse}, \citenamefont {Bender}, \citenamefont {Berti}, \citenamefont {Binetruy}, \citenamefont {Born}, \citenamefont {Bortoluzzi} \emph {et~al.}}]{amaro2017laser}%
  \BibitemOpen
  \bibfield  {author} {\bibinfo {author} {\bibfnamefont {P.}~\bibnamefont {Amaro-Seoane}}, \bibinfo {author} {\bibfnamefont {H.}~\bibnamefont {Audley}}, \bibinfo {author} {\bibfnamefont {S.}~\bibnamefont {Babak}}, \bibinfo {author} {\bibfnamefont {J.}~\bibnamefont {Baker}}, \bibinfo {author} {\bibfnamefont {E.}~\bibnamefont {Barausse}}, \bibinfo {author} {\bibfnamefont {P.}~\bibnamefont {Bender}}, \bibinfo {author} {\bibfnamefont {E.}~\bibnamefont {Berti}}, \bibinfo {author} {\bibfnamefont {P.}~\bibnamefont {Binetruy}}, \bibinfo {author} {\bibfnamefont {M.}~\bibnamefont {Born}}, \bibinfo {author} {\bibfnamefont {D.}~\bibnamefont {Bortoluzzi}}, \emph {et~al.},\ }\href@noop {} {\bibfield  {journal} {\bibinfo  {journal} {arXiv preprint arXiv:1702.00786}\ } (\bibinfo {year} {2017})}\BibitemShut {NoStop}%
\bibitem [{\citenamefont {Hu}\ and\ \citenamefont {Wu}(2017)}]{hu2017taiji}%
  \BibitemOpen
  \bibfield  {author} {\bibinfo {author} {\bibfnamefont {W.-R.}\ \bibnamefont {Hu}}\ and\ \bibinfo {author} {\bibfnamefont {Y.-L.}\ \bibnamefont {Wu}},\ }\href@noop {} {\bibinfo {title} {The taiji program in space for gravitational wave physics and the nature of gravity}} (\bibinfo {year} {2017})\BibitemShut {NoStop}%
\bibitem [{\citenamefont {Luo}\ \emph {et~al.}(2016)\citenamefont {Luo}, \citenamefont {Chen}, \citenamefont {Duan}, \citenamefont {Gong}, \citenamefont {Hu}, \citenamefont {Ji}, \citenamefont {Liu}, \citenamefont {Mei}, \citenamefont {Milyukov}, \citenamefont {Sazhin} \emph {et~al.}}]{luo2016tianqin}%
  \BibitemOpen
  \bibfield  {author} {\bibinfo {author} {\bibfnamefont {J.}~\bibnamefont {Luo}}, \bibinfo {author} {\bibfnamefont {L.-S.}\ \bibnamefont {Chen}}, \bibinfo {author} {\bibfnamefont {H.-Z.}\ \bibnamefont {Duan}}, \bibinfo {author} {\bibfnamefont {Y.-G.}\ \bibnamefont {Gong}}, \bibinfo {author} {\bibfnamefont {S.}~\bibnamefont {Hu}}, \bibinfo {author} {\bibfnamefont {J.}~\bibnamefont {Ji}}, \bibinfo {author} {\bibfnamefont {Q.}~\bibnamefont {Liu}}, \bibinfo {author} {\bibfnamefont {J.}~\bibnamefont {Mei}}, \bibinfo {author} {\bibfnamefont {V.}~\bibnamefont {Milyukov}}, \bibinfo {author} {\bibfnamefont {M.}~\bibnamefont {Sazhin}}, \emph {et~al.},\ }\href@noop {} {\bibfield  {journal} {\bibinfo  {journal} {Classical and Quantum Gravity}\ }\textbf {\bibinfo {volume} {33}},\ \bibinfo {pages} {035010} (\bibinfo {year} {2016})}\BibitemShut {NoStop}%
\bibitem [{\citenamefont {Klein}\ \emph {et~al.}(2016)\citenamefont {Klein}, \citenamefont {Barausse}, \citenamefont {Sesana}, \citenamefont {Petiteau}, \citenamefont {Berti}, \citenamefont {Babak}, \citenamefont {Gair}, \citenamefont {Aoudia}, \citenamefont {Hinder}, \citenamefont {Ohme} \emph {et~al.}}]{klein2016science}%
  \BibitemOpen
  \bibfield  {author} {\bibinfo {author} {\bibfnamefont {A.}~\bibnamefont {Klein}}, \bibinfo {author} {\bibfnamefont {E.}~\bibnamefont {Barausse}}, \bibinfo {author} {\bibfnamefont {A.}~\bibnamefont {Sesana}}, \bibinfo {author} {\bibfnamefont {A.}~\bibnamefont {Petiteau}}, \bibinfo {author} {\bibfnamefont {E.}~\bibnamefont {Berti}}, \bibinfo {author} {\bibfnamefont {S.}~\bibnamefont {Babak}}, \bibinfo {author} {\bibfnamefont {J.}~\bibnamefont {Gair}}, \bibinfo {author} {\bibfnamefont {S.}~\bibnamefont {Aoudia}}, \bibinfo {author} {\bibfnamefont {I.}~\bibnamefont {Hinder}}, \bibinfo {author} {\bibfnamefont {F.}~\bibnamefont {Ohme}}, \emph {et~al.},\ }\href@noop {} {\bibfield  {journal} {\bibinfo  {journal} {Physical Review D}\ }\textbf {\bibinfo {volume} {93}},\ \bibinfo {pages} {024003} (\bibinfo {year} {2016})}\BibitemShut {NoStop}%
\bibitem [{\citenamefont {Owen}\ and\ \citenamefont {Sathyaprakash}(1999)}]{owen1999matched}%
  \BibitemOpen
  \bibfield  {author} {\bibinfo {author} {\bibfnamefont {B.~J.}\ \bibnamefont {Owen}}\ and\ \bibinfo {author} {\bibfnamefont {B.~S.}\ \bibnamefont {Sathyaprakash}},\ }\href@noop {} {\bibfield  {journal} {\bibinfo  {journal} {Physical Review D}\ }\textbf {\bibinfo {volume} {60}},\ \bibinfo {pages} {022002} (\bibinfo {year} {1999})}\BibitemShut {NoStop}%
\bibitem [{\citenamefont {Allen}\ \emph {et~al.}(2012)\citenamefont {Allen}, \citenamefont {Anderson}, \citenamefont {Brady}, \citenamefont {Brown},\ and\ \citenamefont {Creighton}}]{allen2012findchirp}%
  \BibitemOpen
  \bibfield  {author} {\bibinfo {author} {\bibfnamefont {B.}~\bibnamefont {Allen}}, \bibinfo {author} {\bibfnamefont {W.~G.}\ \bibnamefont {Anderson}}, \bibinfo {author} {\bibfnamefont {P.~R.}\ \bibnamefont {Brady}}, \bibinfo {author} {\bibfnamefont {D.~A.}\ \bibnamefont {Brown}},\ and\ \bibinfo {author} {\bibfnamefont {J.~D.}\ \bibnamefont {Creighton}},\ }\href@noop {} {\bibfield  {journal} {\bibinfo  {journal} {Physical Review D}\ }\textbf {\bibinfo {volume} {85}},\ \bibinfo {pages} {122006} (\bibinfo {year} {2012})}\BibitemShut {NoStop}%
\bibitem [{\citenamefont {Abbott}\ \emph {et~al.}(2016{\natexlab{a}})\citenamefont {Abbott}, \citenamefont {Jawahar}, \citenamefont {Lockerbie},\ and\ \citenamefont {Tokmakov}}]{abbott2016ligo}%
  \BibitemOpen
  \bibfield  {author} {\bibinfo {author} {\bibfnamefont {B.}~\bibnamefont {Abbott}}, \bibinfo {author} {\bibfnamefont {S.}~\bibnamefont {Jawahar}}, \bibinfo {author} {\bibfnamefont {N.}~\bibnamefont {Lockerbie}},\ and\ \bibinfo {author} {\bibfnamefont {K.}~\bibnamefont {Tokmakov}},\ }\href@noop {} {\bibfield  {journal} {\bibinfo  {journal} {PHYSICAL REVIEW D Phys Rev D}\ }\textbf {\bibinfo {volume} {93}},\ \bibinfo {pages} {122003} (\bibinfo {year} {2016}{\natexlab{a}})}\BibitemShut {NoStop}%
\bibitem [{\citenamefont {Abbott}\ \emph {et~al.}(2016{\natexlab{b}})\citenamefont {Abbott}, \citenamefont {Abbott}, \citenamefont {Abbott}, \citenamefont {Abernathy}, \citenamefont {Acernese}, \citenamefont {Ackley}, \citenamefont {Adams}, \citenamefont {Adams}, \citenamefont {Addesso}, \citenamefont {Adhikari} \emph {et~al.}}]{abbott2016gw151226}%
  \BibitemOpen
  \bibfield  {author} {\bibinfo {author} {\bibfnamefont {B.~P.}\ \bibnamefont {Abbott}}, \bibinfo {author} {\bibfnamefont {R.}~\bibnamefont {Abbott}}, \bibinfo {author} {\bibfnamefont {T.}~\bibnamefont {Abbott}}, \bibinfo {author} {\bibfnamefont {M.}~\bibnamefont {Abernathy}}, \bibinfo {author} {\bibfnamefont {F.}~\bibnamefont {Acernese}}, \bibinfo {author} {\bibfnamefont {K.}~\bibnamefont {Ackley}}, \bibinfo {author} {\bibfnamefont {C.}~\bibnamefont {Adams}}, \bibinfo {author} {\bibfnamefont {T.}~\bibnamefont {Adams}}, \bibinfo {author} {\bibfnamefont {P.}~\bibnamefont {Addesso}}, \bibinfo {author} {\bibfnamefont {R.}~\bibnamefont {Adhikari}}, \emph {et~al.},\ }\href@noop {} {\bibfield  {journal} {\bibinfo  {journal} {Physical review letters}\ }\textbf {\bibinfo {volume} {116}},\ \bibinfo {pages} {241103} (\bibinfo {year} {2016}{\natexlab{b}})}\BibitemShut {NoStop}%
\bibitem [{\citenamefont {Tinto}\ and\ \citenamefont {Dhurandhar}(2014)}]{tinto2014time}%
  \BibitemOpen
  \bibfield  {author} {\bibinfo {author} {\bibfnamefont {M.}~\bibnamefont {Tinto}}\ and\ \bibinfo {author} {\bibfnamefont {S.~V.}\ \bibnamefont {Dhurandhar}},\ }\href@noop {} {\bibfield  {journal} {\bibinfo  {journal} {Living Reviews in Relativity}\ }\textbf {\bibinfo {volume} {17}},\ \bibinfo {pages} {1} (\bibinfo {year} {2014})}\BibitemShut {NoStop}%
\bibitem [{\citenamefont {Cornish}\ and\ \citenamefont {Crowder}(2005)}]{cornish2005lisa}%
  \BibitemOpen
  \bibfield  {author} {\bibinfo {author} {\bibfnamefont {N.~J.}\ \bibnamefont {Cornish}}\ and\ \bibinfo {author} {\bibfnamefont {J.}~\bibnamefont {Crowder}},\ }\href@noop {} {\bibfield  {journal} {\bibinfo  {journal} {Physical Review D}\ }\textbf {\bibinfo {volume} {72}},\ \bibinfo {pages} {043005} (\bibinfo {year} {2005})}\BibitemShut {NoStop}%
\bibitem [{\citenamefont {Littenberg}\ \emph {et~al.}(2020)\citenamefont {Littenberg}, \citenamefont {Cornish}, \citenamefont {Lackeos},\ and\ \citenamefont {Robson}}]{littenberg2020global}%
  \BibitemOpen
  \bibfield  {author} {\bibinfo {author} {\bibfnamefont {T.~B.}\ \bibnamefont {Littenberg}}, \bibinfo {author} {\bibfnamefont {N.~J.}\ \bibnamefont {Cornish}}, \bibinfo {author} {\bibfnamefont {K.}~\bibnamefont {Lackeos}},\ and\ \bibinfo {author} {\bibfnamefont {T.}~\bibnamefont {Robson}},\ }\href@noop {} {\bibfield  {journal} {\bibinfo  {journal} {Physical Review D}\ }\textbf {\bibinfo {volume} {101}},\ \bibinfo {pages} {123021} (\bibinfo {year} {2020})}\BibitemShut {NoStop}%
\bibitem [{\citenamefont {Littenberg}\ and\ \citenamefont {Cornish}(2023)}]{littenberg2023prototype}%
  \BibitemOpen
  \bibfield  {author} {\bibinfo {author} {\bibfnamefont {T.~B.}\ \bibnamefont {Littenberg}}\ and\ \bibinfo {author} {\bibfnamefont {N.~J.}\ \bibnamefont {Cornish}},\ }\href@noop {} {\bibfield  {journal} {\bibinfo  {journal} {Physical Review D}\ }\textbf {\bibinfo {volume} {107}},\ \bibinfo {pages} {063004} (\bibinfo {year} {2023})}\BibitemShut {NoStop}%
\bibitem [{\citenamefont {George}\ and\ \citenamefont {Huerta}(2018)}]{george2018deep}%
  \BibitemOpen
  \bibfield  {author} {\bibinfo {author} {\bibfnamefont {D.}~\bibnamefont {George}}\ and\ \bibinfo {author} {\bibfnamefont {E.~A.}\ \bibnamefont {Huerta}},\ }\href@noop {} {\bibfield  {journal} {\bibinfo  {journal} {Physics Letters B}\ }\textbf {\bibinfo {volume} {778}},\ \bibinfo {pages} {64} (\bibinfo {year} {2018})}\BibitemShut {NoStop}%
\bibitem [{\citenamefont {Green}\ \emph {et~al.}(2020)\citenamefont {Green}, \citenamefont {Simpson},\ and\ \citenamefont {Gair}}]{green2020gravitational}%
  \BibitemOpen
  \bibfield  {author} {\bibinfo {author} {\bibfnamefont {S.~R.}\ \bibnamefont {Green}}, \bibinfo {author} {\bibfnamefont {C.}~\bibnamefont {Simpson}},\ and\ \bibinfo {author} {\bibfnamefont {J.}~\bibnamefont {Gair}},\ }\href@noop {} {\bibfield  {journal} {\bibinfo  {journal} {Physical Review D}\ }\textbf {\bibinfo {volume} {102}},\ \bibinfo {pages} {104057} (\bibinfo {year} {2020})}\BibitemShut {NoStop}%
\bibitem [{\citenamefont {Krastev}\ \emph {et~al.}(2021)\citenamefont {Krastev}, \citenamefont {Gill}, \citenamefont {Villar},\ and\ \citenamefont {Berger}}]{krastev2021detection}%
  \BibitemOpen
  \bibfield  {author} {\bibinfo {author} {\bibfnamefont {P.~G.}\ \bibnamefont {Krastev}}, \bibinfo {author} {\bibfnamefont {K.}~\bibnamefont {Gill}}, \bibinfo {author} {\bibfnamefont {V.~A.}\ \bibnamefont {Villar}},\ and\ \bibinfo {author} {\bibfnamefont {E.}~\bibnamefont {Berger}},\ }\href@noop {} {\bibfield  {journal} {\bibinfo  {journal} {Physics Letters B}\ }\textbf {\bibinfo {volume} {815}},\ \bibinfo {pages} {136161} (\bibinfo {year} {2021})}\BibitemShut {NoStop}%
\bibitem [{\citenamefont {Green}\ and\ \citenamefont {Gair}(2021)}]{green2021complete}%
  \BibitemOpen
  \bibfield  {author} {\bibinfo {author} {\bibfnamefont {S.~R.}\ \bibnamefont {Green}}\ and\ \bibinfo {author} {\bibfnamefont {J.}~\bibnamefont {Gair}},\ }\href@noop {} {\bibfield  {journal} {\bibinfo  {journal} {Machine Learning: Science and Technology}\ }\textbf {\bibinfo {volume} {2}},\ \bibinfo {pages} {03LT01} (\bibinfo {year} {2021})}\BibitemShut {NoStop}%
\bibitem [{\citenamefont {Dax}\ \emph {et~al.}(2021)\citenamefont {Dax}, \citenamefont {Green}, \citenamefont {Gair}, \citenamefont {Macke}, \citenamefont {Buonanno},\ and\ \citenamefont {Sch{\"o}lkopf}}]{dax2021real}%
  \BibitemOpen
  \bibfield  {author} {\bibinfo {author} {\bibfnamefont {M.}~\bibnamefont {Dax}}, \bibinfo {author} {\bibfnamefont {S.~R.}\ \bibnamefont {Green}}, \bibinfo {author} {\bibfnamefont {J.}~\bibnamefont {Gair}}, \bibinfo {author} {\bibfnamefont {J.~H.}\ \bibnamefont {Macke}}, \bibinfo {author} {\bibfnamefont {A.}~\bibnamefont {Buonanno}},\ and\ \bibinfo {author} {\bibfnamefont {B.}~\bibnamefont {Sch{\"o}lkopf}},\ }\href@noop {} {\bibfield  {journal} {\bibinfo  {journal} {Physical review letters}\ }\textbf {\bibinfo {volume} {127}},\ \bibinfo {pages} {241103} (\bibinfo {year} {2021})}\BibitemShut {NoStop}%
\bibitem [{\citenamefont {Shen}\ \emph {et~al.}(2021)\citenamefont {Shen}, \citenamefont {Huerta}, \citenamefont {O’Shea}, \citenamefont {Kumar},\ and\ \citenamefont {Zhao}}]{shen2021statistically}%
  \BibitemOpen
  \bibfield  {author} {\bibinfo {author} {\bibfnamefont {H.}~\bibnamefont {Shen}}, \bibinfo {author} {\bibfnamefont {E.}~\bibnamefont {Huerta}}, \bibinfo {author} {\bibfnamefont {E.}~\bibnamefont {O’Shea}}, \bibinfo {author} {\bibfnamefont {P.}~\bibnamefont {Kumar}},\ and\ \bibinfo {author} {\bibfnamefont {Z.}~\bibnamefont {Zhao}},\ }\href@noop {} {\bibfield  {journal} {\bibinfo  {journal} {Machine Learning: Science and Technology}\ }\textbf {\bibinfo {volume} {3}},\ \bibinfo {pages} {015007} (\bibinfo {year} {2021})}\BibitemShut {NoStop}%
\bibitem [{\citenamefont {Schmidt}\ \emph {et~al.}(2021)\citenamefont {Schmidt}, \citenamefont {Breschi}, \citenamefont {Gamba}, \citenamefont {Pagano}, \citenamefont {Rettegno}, \citenamefont {Riemenschneider}, \citenamefont {Bernuzzi}, \citenamefont {Nagar},\ and\ \citenamefont {Del~Pozzo}}]{schmidt2021machine}%
  \BibitemOpen
  \bibfield  {author} {\bibinfo {author} {\bibfnamefont {S.}~\bibnamefont {Schmidt}}, \bibinfo {author} {\bibfnamefont {M.}~\bibnamefont {Breschi}}, \bibinfo {author} {\bibfnamefont {R.}~\bibnamefont {Gamba}}, \bibinfo {author} {\bibfnamefont {G.}~\bibnamefont {Pagano}}, \bibinfo {author} {\bibfnamefont {P.}~\bibnamefont {Rettegno}}, \bibinfo {author} {\bibfnamefont {G.}~\bibnamefont {Riemenschneider}}, \bibinfo {author} {\bibfnamefont {S.}~\bibnamefont {Bernuzzi}}, \bibinfo {author} {\bibfnamefont {A.}~\bibnamefont {Nagar}},\ and\ \bibinfo {author} {\bibfnamefont {W.}~\bibnamefont {Del~Pozzo}},\ }\href@noop {} {\bibfield  {journal} {\bibinfo  {journal} {Physical Review D}\ }\textbf {\bibinfo {volume} {103}},\ \bibinfo {pages} {043020} (\bibinfo {year} {2021})}\BibitemShut {NoStop}%
\bibitem [{\citenamefont {Gabbard}\ \emph {et~al.}(2022)\citenamefont {Gabbard}, \citenamefont {Messenger}, \citenamefont {Heng}, \citenamefont {Tonolini},\ and\ \citenamefont {Murray-Smith}}]{gabbard2022bayesian}%
  \BibitemOpen
  \bibfield  {author} {\bibinfo {author} {\bibfnamefont {H.}~\bibnamefont {Gabbard}}, \bibinfo {author} {\bibfnamefont {C.}~\bibnamefont {Messenger}}, \bibinfo {author} {\bibfnamefont {I.~S.}\ \bibnamefont {Heng}}, \bibinfo {author} {\bibfnamefont {F.}~\bibnamefont {Tonolini}},\ and\ \bibinfo {author} {\bibfnamefont {R.}~\bibnamefont {Murray-Smith}},\ }\href@noop {} {\bibfield  {journal} {\bibinfo  {journal} {Nature Physics}\ }\textbf {\bibinfo {volume} {18}},\ \bibinfo {pages} {112} (\bibinfo {year} {2022})}\BibitemShut {NoStop}%
\bibitem [{\citenamefont {Langendorff}\ \emph {et~al.}(2023)\citenamefont {Langendorff}, \citenamefont {Kolmus}, \citenamefont {Janquart},\ and\ \citenamefont {Van Den~Broeck}}]{langendorff2023normalizing}%
  \BibitemOpen
  \bibfield  {author} {\bibinfo {author} {\bibfnamefont {J.}~\bibnamefont {Langendorff}}, \bibinfo {author} {\bibfnamefont {A.}~\bibnamefont {Kolmus}}, \bibinfo {author} {\bibfnamefont {J.}~\bibnamefont {Janquart}},\ and\ \bibinfo {author} {\bibfnamefont {C.}~\bibnamefont {Van Den~Broeck}},\ }\href@noop {} {\bibfield  {journal} {\bibinfo  {journal} {Physical Review Letters}\ }\textbf {\bibinfo {volume} {130}},\ \bibinfo {pages} {171402} (\bibinfo {year} {2023})}\BibitemShut {NoStop}%
\bibitem [{\citenamefont {Chua}\ and\ \citenamefont {Vallisneri}(2020)}]{chua2020learning}%
  \BibitemOpen
  \bibfield  {author} {\bibinfo {author} {\bibfnamefont {A.~J.}\ \bibnamefont {Chua}}\ and\ \bibinfo {author} {\bibfnamefont {M.}~\bibnamefont {Vallisneri}},\ }\href@noop {} {\bibfield  {journal} {\bibinfo  {journal} {Physical review letters}\ }\textbf {\bibinfo {volume} {124}},\ \bibinfo {pages} {041102} (\bibinfo {year} {2020})}\BibitemShut {NoStop}%
\bibitem [{\citenamefont {Rezende}\ and\ \citenamefont {Mohamed}(2015)}]{rezende2015variational}%
  \BibitemOpen
  \bibfield  {author} {\bibinfo {author} {\bibfnamefont {D.}~\bibnamefont {Rezende}}\ and\ \bibinfo {author} {\bibfnamefont {S.}~\bibnamefont {Mohamed}},\ }in\ \href@noop {} {\emph {\bibinfo {booktitle} {International conference on machine learning}}}\ (\bibinfo {organization} {PMLR},\ \bibinfo {year} {2015})\ pp.\ \bibinfo {pages} {1530--1538}\BibitemShut {NoStop}%
\bibitem [{\citenamefont {LeCun}\ \emph {et~al.}(1998)\citenamefont {LeCun}, \citenamefont {Bottou}, \citenamefont {Bengio},\ and\ \citenamefont {Haffner}}]{lecun1998gradient}%
  \BibitemOpen
  \bibfield  {author} {\bibinfo {author} {\bibfnamefont {Y.}~\bibnamefont {LeCun}}, \bibinfo {author} {\bibfnamefont {L.}~\bibnamefont {Bottou}}, \bibinfo {author} {\bibfnamefont {Y.}~\bibnamefont {Bengio}},\ and\ \bibinfo {author} {\bibfnamefont {P.}~\bibnamefont {Haffner}},\ }\href@noop {} {\bibfield  {journal} {\bibinfo  {journal} {Proceedings of the IEEE}\ }\textbf {\bibinfo {volume} {86}},\ \bibinfo {pages} {2278} (\bibinfo {year} {1998})}\BibitemShut {NoStop}%
\bibitem [{\citenamefont {Durkan}\ \emph {et~al.}(2019)\citenamefont {Durkan}, \citenamefont {Bekasov}, \citenamefont {Murray},\ and\ \citenamefont {Papamakarios}}]{durkan2019neural}%
  \BibitemOpen
  \bibfield  {author} {\bibinfo {author} {\bibfnamefont {C.}~\bibnamefont {Durkan}}, \bibinfo {author} {\bibfnamefont {A.}~\bibnamefont {Bekasov}}, \bibinfo {author} {\bibfnamefont {I.}~\bibnamefont {Murray}},\ and\ \bibinfo {author} {\bibfnamefont {G.}~\bibnamefont {Papamakarios}},\ }\href@noop {} {\bibfield  {journal} {\bibinfo  {journal} {Advances in neural information processing systems}\ }\textbf {\bibinfo {volume} {32}} (\bibinfo {year} {2019})}\BibitemShut {NoStop}%
\bibitem [{\citenamefont {Oliva}\ \emph {et~al.}(2018)\citenamefont {Oliva}, \citenamefont {Dubey}, \citenamefont {Zaheer}, \citenamefont {Poczos}, \citenamefont {Salakhutdinov}, \citenamefont {Xing},\ and\ \citenamefont {Schneider}}]{oliva2018transformation}%
  \BibitemOpen
  \bibfield  {author} {\bibinfo {author} {\bibfnamefont {J.}~\bibnamefont {Oliva}}, \bibinfo {author} {\bibfnamefont {A.}~\bibnamefont {Dubey}}, \bibinfo {author} {\bibfnamefont {M.}~\bibnamefont {Zaheer}}, \bibinfo {author} {\bibfnamefont {B.}~\bibnamefont {Poczos}}, \bibinfo {author} {\bibfnamefont {R.}~\bibnamefont {Salakhutdinov}}, \bibinfo {author} {\bibfnamefont {E.}~\bibnamefont {Xing}},\ and\ \bibinfo {author} {\bibfnamefont {J.}~\bibnamefont {Schneider}},\ }in\ \href@noop {} {\emph {\bibinfo {booktitle} {International Conference on Machine Learning}}}\ (\bibinfo {organization} {PMLR},\ \bibinfo {year} {2018})\ pp.\ \bibinfo {pages} {3898--3907}\BibitemShut {NoStop}%
\bibitem [{\citenamefont {He}\ \emph {et~al.}(2016)\citenamefont {He}, \citenamefont {Zhang}, \citenamefont {Ren},\ and\ \citenamefont {Sun}}]{he2016deep}%
  \BibitemOpen
  \bibfield  {author} {\bibinfo {author} {\bibfnamefont {K.}~\bibnamefont {He}}, \bibinfo {author} {\bibfnamefont {X.}~\bibnamefont {Zhang}}, \bibinfo {author} {\bibfnamefont {S.}~\bibnamefont {Ren}},\ and\ \bibinfo {author} {\bibfnamefont {J.}~\bibnamefont {Sun}},\ }in\ \href@noop {} {\emph {\bibinfo {booktitle} {Proceedings of the IEEE conference on computer vision and pattern recognition}}}\ (\bibinfo {year} {2016})\ pp.\ \bibinfo {pages} {770--778}\BibitemShut {NoStop}%
\bibitem [{\citenamefont {Kingma}\ and\ \citenamefont {Ba}(2014)}]{kingma2014adam}%
  \BibitemOpen
  \bibfield  {author} {\bibinfo {author} {\bibfnamefont {D.~P.}\ \bibnamefont {Kingma}}\ and\ \bibinfo {author} {\bibfnamefont {J.}~\bibnamefont {Ba}},\ }\href@noop {} {\bibfield  {journal} {\bibinfo  {journal} {arXiv preprint arXiv:1412.6980}\ } (\bibinfo {year} {2014})}\BibitemShut {NoStop}%
\bibitem [{\citenamefont {Paszke}\ \emph {et~al.}(2019)\citenamefont {Paszke}, \citenamefont {Gross}, \citenamefont {Massa}, \citenamefont {Lerer}, \citenamefont {Bradbury}, \citenamefont {Chanan}, \citenamefont {Killeen}, \citenamefont {Lin}, \citenamefont {Gimelshein}, \citenamefont {Antiga}, \citenamefont {Desmaison}, \citenamefont {Kopf}, \citenamefont {Yang}, \citenamefont {DeVito}, \citenamefont {Raison}, \citenamefont {Tejani}, \citenamefont {Chilamkurthy}, \citenamefont {Steiner}, \citenamefont {Fang}, \citenamefont {Bai},\ and\ \citenamefont {Chintala}}]{NEURIPS2019_9015}%
  \BibitemOpen
  \bibfield  {author} {\bibinfo {author} {\bibfnamefont {A.}~\bibnamefont {Paszke}}, \bibinfo {author} {\bibfnamefont {S.}~\bibnamefont {Gross}}, \bibinfo {author} {\bibfnamefont {F.}~\bibnamefont {Massa}}, \bibinfo {author} {\bibfnamefont {A.}~\bibnamefont {Lerer}}, \bibinfo {author} {\bibfnamefont {J.}~\bibnamefont {Bradbury}}, \bibinfo {author} {\bibfnamefont {G.}~\bibnamefont {Chanan}}, \bibinfo {author} {\bibfnamefont {T.}~\bibnamefont {Killeen}}, \bibinfo {author} {\bibfnamefont {Z.}~\bibnamefont {Lin}}, \bibinfo {author} {\bibfnamefont {N.}~\bibnamefont {Gimelshein}}, \bibinfo {author} {\bibfnamefont {L.}~\bibnamefont {Antiga}}, \bibinfo {author} {\bibfnamefont {A.}~\bibnamefont {Desmaison}}, \bibinfo {author} {\bibfnamefont {A.}~\bibnamefont {Kopf}}, \bibinfo {author} {\bibfnamefont {E.}~\bibnamefont {Yang}}, \bibinfo {author} {\bibfnamefont {Z.}~\bibnamefont {DeVito}}, \bibinfo {author} {\bibfnamefont {M.}~\bibnamefont {Raison}}, \bibinfo {author} {\bibfnamefont {A.}~\bibnamefont {Tejani}}, \bibinfo {author} {\bibfnamefont {S.}~\bibnamefont {Chilamkurthy}}, \bibinfo {author} {\bibfnamefont {B.}~\bibnamefont {Steiner}}, \bibinfo {author} {\bibfnamefont {L.}~\bibnamefont {Fang}}, \bibinfo {author} {\bibfnamefont {J.}~\bibnamefont {Bai}},\ and\ \bibinfo {author} {\bibfnamefont {S.}~\bibnamefont {Chintala}},\ }in\ \href {http://papers.neurips.cc/paper/9015-pytorch-an-imperative-style-high-performance-deep-learning-library.pdf} {\emph {\bibinfo {booktitle} {Advances in Neural Information Processing Systems 32}}},\ \bibinfo {editor} {edited by\ \bibinfo {editor} {\bibfnamefont {H.}~\bibnamefont {Wallach}}, \bibinfo {editor} {\bibfnamefont {H.}~\bibnamefont {Larochelle}}, \bibinfo {editor} {\bibfnamefont {A.}~\bibnamefont {Beygelzimer}}, \bibinfo {editor} {\bibfnamefont {F.}~\bibnamefont {d~Alche-Buc}}, \bibinfo {editor} {\bibfnamefont {E.}~\bibnamefont {Fox}},\ and\ \bibinfo {editor} {\bibfnamefont {R.}~\bibnamefont {Garnett}}}\ (\bibinfo  {publisher} {Curran Associates, Inc.},\ \bibinfo {year} {2019})\ pp.\ \bibinfo {pages} {8024--8035}\BibitemShut {NoStop}%
\bibitem [{\citenamefont {Durkan}\ \emph {et~al.}(2020)\citenamefont {Durkan}, \citenamefont {Bekasov}, \citenamefont {Murray},\ and\ \citenamefont {Papamakarios}}]{nflows}%
  \BibitemOpen
  \bibfield  {author} {\bibinfo {author} {\bibfnamefont {C.}~\bibnamefont {Durkan}}, \bibinfo {author} {\bibfnamefont {A.}~\bibnamefont {Bekasov}}, \bibinfo {author} {\bibfnamefont {I.}~\bibnamefont {Murray}},\ and\ \bibinfo {author} {\bibfnamefont {G.}~\bibnamefont {Papamakarios}},\ }\href {https://doi.org/10.5281/zenodo.4296287} {\bibinfo {title} {{nflows}: normalizing flows in {PyTorch}}} (\bibinfo {year} {2020})\BibitemShut {NoStop}%
\bibitem [{lfi()}]{lfigw}%
  \BibitemOpen
  \href@noop {} {\bibinfo {title} {{lfigw: Likelihood-Free Inference for Gravitational Waves}}},\ \bibinfo {howpublished} {\url{https://github.com/stephengreen/lfi-gw}}\BibitemShut {NoStop}%
\bibitem [{\citenamefont {Husa}\ \emph {et~al.}(2016)\citenamefont {Husa}, \citenamefont {Khan}, \citenamefont {Hannam}, \citenamefont {P{\"u}rrer}, \citenamefont {Ohme}, \citenamefont {Forteza},\ and\ \citenamefont {Boh{\'e}}}]{husa2016frequency}%
  \BibitemOpen
  \bibfield  {author} {\bibinfo {author} {\bibfnamefont {S.}~\bibnamefont {Husa}}, \bibinfo {author} {\bibfnamefont {S.}~\bibnamefont {Khan}}, \bibinfo {author} {\bibfnamefont {M.}~\bibnamefont {Hannam}}, \bibinfo {author} {\bibfnamefont {M.}~\bibnamefont {P{\"u}rrer}}, \bibinfo {author} {\bibfnamefont {F.}~\bibnamefont {Ohme}}, \bibinfo {author} {\bibfnamefont {X.~J.}\ \bibnamefont {Forteza}},\ and\ \bibinfo {author} {\bibfnamefont {A.}~\bibnamefont {Boh{\'e}}},\ }\href@noop {} {\bibfield  {journal} {\bibinfo  {journal} {Physical Review D}\ }\textbf {\bibinfo {volume} {93}},\ \bibinfo {pages} {044006} (\bibinfo {year} {2016})}\BibitemShut {NoStop}%
\bibitem [{\citenamefont {Khan}\ \emph {et~al.}(2016)\citenamefont {Khan}, \citenamefont {Husa}, \citenamefont {Hannam}, \citenamefont {Ohme}, \citenamefont {P{\"u}rrer}, \citenamefont {Forteza},\ and\ \citenamefont {Boh{\'e}}}]{khan2016frequency}%
  \BibitemOpen
  \bibfield  {author} {\bibinfo {author} {\bibfnamefont {S.}~\bibnamefont {Khan}}, \bibinfo {author} {\bibfnamefont {S.}~\bibnamefont {Husa}}, \bibinfo {author} {\bibfnamefont {M.}~\bibnamefont {Hannam}}, \bibinfo {author} {\bibfnamefont {F.}~\bibnamefont {Ohme}}, \bibinfo {author} {\bibfnamefont {M.}~\bibnamefont {P{\"u}rrer}}, \bibinfo {author} {\bibfnamefont {X.~J.}\ \bibnamefont {Forteza}},\ and\ \bibinfo {author} {\bibfnamefont {A.}~\bibnamefont {Boh{\'e}}},\ }\href@noop {} {\bibfield  {journal} {\bibinfo  {journal} {Physical Review D}\ }\textbf {\bibinfo {volume} {93}},\ \bibinfo {pages} {044007} (\bibinfo {year} {2016})}\BibitemShut {NoStop}%
\bibitem [{\citenamefont {Ade}\ \emph {et~al.}(2016)\citenamefont {Ade}, \citenamefont {Aghanim}, \citenamefont {Arnaud}, \citenamefont {Ashdown}, \citenamefont {Aumont}, \citenamefont {Baccigalupi}, \citenamefont {Banday}, \citenamefont {Barreiro}, \citenamefont {Bartlett}, \citenamefont {Bartolo} \emph {et~al.}}]{ade2016planck}%
  \BibitemOpen
  \bibfield  {author} {\bibinfo {author} {\bibfnamefont {P.~A.}\ \bibnamefont {Ade}}, \bibinfo {author} {\bibfnamefont {N.}~\bibnamefont {Aghanim}}, \bibinfo {author} {\bibfnamefont {M.}~\bibnamefont {Arnaud}}, \bibinfo {author} {\bibfnamefont {M.}~\bibnamefont {Ashdown}}, \bibinfo {author} {\bibfnamefont {J.}~\bibnamefont {Aumont}}, \bibinfo {author} {\bibfnamefont {C.}~\bibnamefont {Baccigalupi}}, \bibinfo {author} {\bibfnamefont {A.}~\bibnamefont {Banday}}, \bibinfo {author} {\bibfnamefont {R.}~\bibnamefont {Barreiro}}, \bibinfo {author} {\bibfnamefont {J.}~\bibnamefont {Bartlett}}, \bibinfo {author} {\bibfnamefont {N.}~\bibnamefont {Bartolo}}, \emph {et~al.},\ }\href@noop {} {\bibfield  {journal} {\bibinfo  {journal} {Astronomy \& Astrophysics}\ }\textbf {\bibinfo {volume} {594}},\ \bibinfo {pages} {A13} (\bibinfo {year} {2016})}\BibitemShut {NoStop}%
\bibitem [{\citenamefont {Prince}\ \emph {et~al.}(2002)\citenamefont {Prince}, \citenamefont {Tinto}, \citenamefont {Larson},\ and\ \citenamefont {Armstrong}}]{prince2002lisa}%
  \BibitemOpen
  \bibfield  {author} {\bibinfo {author} {\bibfnamefont {T.~A.}\ \bibnamefont {Prince}}, \bibinfo {author} {\bibfnamefont {M.}~\bibnamefont {Tinto}}, \bibinfo {author} {\bibfnamefont {S.~L.}\ \bibnamefont {Larson}},\ and\ \bibinfo {author} {\bibfnamefont {J.}~\bibnamefont {Armstrong}},\ }\href@noop {} {\bibfield  {journal} {\bibinfo  {journal} {Physical Review D}\ }\textbf {\bibinfo {volume} {66}},\ \bibinfo {pages} {122002} (\bibinfo {year} {2002})}\BibitemShut {NoStop}%
\bibitem [{\citenamefont {{The LISA Science Study Team}}(2018)}]{lisa_sci_rs}%
  \BibitemOpen
  \bibfield  {author} {\bibinfo {author} {\bibnamefont {{The LISA Science Study Team}}},\ }\href@noop {} {\bibinfo {title} {{ESA-L3-EST-SCI-RS-001}}},\ \bibinfo {howpublished} {\url{https://atrium.in2p3.fr/f5a78d3e-9e19-47a5-aa11-51c81d370f5f}} (\bibinfo {year} {2018})\BibitemShut {NoStop}%
\bibitem [{\citenamefont {Babak}\ and\ \citenamefont {Petiteau}(2018)}]{ldcmanual001}%
  \BibitemOpen
  \bibfield  {author} {\bibinfo {author} {\bibfnamefont {S.}~\bibnamefont {Babak}}\ and\ \bibinfo {author} {\bibfnamefont {A.}~\bibnamefont {Petiteau}},\ }\href@noop {} {\bibinfo {title} {{LISA Data Challenge Manual}}},\ \bibinfo {howpublished} {\url{https://lisa-ldc.lal.in2p3.fr/static/data/pdf/LDC-manual-002.pdf}} (\bibinfo {year} {2018})\BibitemShut {NoStop}%
\bibitem [{\citenamefont {Nitz}\ \emph {et~al.}(2021)\citenamefont {Nitz}, \citenamefont {Harry}, \citenamefont {Brown}, \citenamefont {Biwer}, \citenamefont {Willis}, \citenamefont {Canton}, \citenamefont {Capano}, \citenamefont {Dent}, \citenamefont {Pekowsky}, \citenamefont {Williamson}, \citenamefont {Davies}, \citenamefont {De}, \citenamefont {Cabero}, \citenamefont {Machenschalk}, \citenamefont {Kumar}, \citenamefont {Macleod}, \citenamefont {Reyes}, \citenamefont {dfinstad}, \citenamefont {Pannarale}, \citenamefont {Massinger}, \citenamefont {Kumar}, \citenamefont {TÃ¡pai}, \citenamefont {Singer}, \citenamefont {Khan}, \citenamefont {Fairhurst}, \citenamefont {Nielsen}, \citenamefont {Singh}, \citenamefont {Chandra}, \citenamefont {shasvath},\ and\ \citenamefont {Gadre}}]{alex_nitz_2021_5347736}%
  \BibitemOpen
  \bibfield  {author} {\bibinfo {author} {\bibfnamefont {A.}~\bibnamefont {Nitz}}, \bibinfo {author} {\bibfnamefont {I.}~\bibnamefont {Harry}}, \bibinfo {author} {\bibfnamefont {D.}~\bibnamefont {Brown}}, \bibinfo {author} {\bibfnamefont {C.~M.}\ \bibnamefont {Biwer}}, \bibinfo {author} {\bibfnamefont {J.}~\bibnamefont {Willis}}, \bibinfo {author} {\bibfnamefont {T.~D.}\ \bibnamefont {Canton}}, \bibinfo {author} {\bibfnamefont {C.}~\bibnamefont {Capano}}, \bibinfo {author} {\bibfnamefont {T.}~\bibnamefont {Dent}}, \bibinfo {author} {\bibfnamefont {L.}~\bibnamefont {Pekowsky}}, \bibinfo {author} {\bibfnamefont {A.~R.}\ \bibnamefont {Williamson}}, \bibinfo {author} {\bibfnamefont {G.~S.~C.}\ \bibnamefont {Davies}}, \bibinfo {author} {\bibfnamefont {S.}~\bibnamefont {De}}, \bibinfo {author} {\bibfnamefont {M.}~\bibnamefont {Cabero}}, \bibinfo {author} {\bibfnamefont {B.}~\bibnamefont {Machenschalk}}, \bibinfo {author} {\bibfnamefont {P.}~\bibnamefont {Kumar}}, \bibinfo {author} {\bibfnamefont {D.}~\bibnamefont {Macleod}}, \bibinfo {author} {\bibfnamefont {S.}~\bibnamefont {Reyes}}, \bibinfo {author} {\bibnamefont {dfinstad}}, \bibinfo {author} {\bibfnamefont {F.}~\bibnamefont {Pannarale}}, \bibinfo {author} {\bibfnamefont {T.}~\bibnamefont {Massinger}}, \bibinfo {author} {\bibfnamefont {S.}~\bibnamefont {Kumar}}, \bibinfo {author} {\bibfnamefont {M.}~\bibnamefont {TÃ¡pai}}, \bibinfo {author} {\bibfnamefont {L.}~\bibnamefont {Singer}}, \bibinfo {author} {\bibfnamefont {S.}~\bibnamefont {Khan}}, \bibinfo {author} {\bibfnamefont {S.}~\bibnamefont {Fairhurst}}, \bibinfo {author} {\bibfnamefont {A.}~\bibnamefont {Nielsen}}, \bibinfo {author} {\bibfnamefont {S.}~\bibnamefont {Singh}}, \bibinfo {author} {\bibfnamefont {K.}~\bibnamefont {Chandra}}, \bibinfo {author} {\bibnamefont {shasvath}},\ and\ \bibinfo {author} {\bibfnamefont {B.~U.~V.}\ \bibnamefont {Gadre}},\ }\href {https://doi.org/10.5281/zenodo.5347736} {\bibinfo {title} {gwastro/pycbc:}} (\bibinfo {year} {2021})\BibitemShut {NoStop}%
\bibitem [{\citenamefont {Loshchilov}\ and\ \citenamefont {Hutter}(2016)}]{loshchilov2016sgdr}%
  \BibitemOpen
  \bibfield  {author} {\bibinfo {author} {\bibfnamefont {I.}~\bibnamefont {Loshchilov}}\ and\ \bibinfo {author} {\bibfnamefont {F.}~\bibnamefont {Hutter}},\ }\href@noop {} {\bibfield  {journal} {\bibinfo  {journal} {arXiv preprint arXiv:1608.03983}\ } (\bibinfo {year} {2016})}\BibitemShut {NoStop}%
\bibitem [{\citenamefont {Massey~Jr}(1951)}]{massey1951kolmogorov}%
  \BibitemOpen
  \bibfield  {author} {\bibinfo {author} {\bibfnamefont {F.~J.}\ \bibnamefont {Massey~Jr}},\ }\href@noop {} {\bibfield  {journal} {\bibinfo  {journal} {Journal of the American statistical Association}\ }\textbf {\bibinfo {volume} {46}},\ \bibinfo {pages} {68} (\bibinfo {year} {1951})}\BibitemShut {NoStop}%
\bibitem [{\citenamefont {{LISA Consortium's LDC working group}}(2019)}]{ldcweb}%
  \BibitemOpen
  \bibfield  {author} {\bibinfo {author} {\bibnamefont {{LISA Consortium's LDC working group}}},\ }\href@noop {} {\bibinfo {title} {{LISA Data Challenges}}},\ \bibinfo {howpublished} {\url{https://lisa-ldc.lal.in2p3.fr}} (\bibinfo {year} {2019})\BibitemShut {NoStop}%
\bibitem [{\citenamefont {Cornish}(2022)}]{cornish2022low}%
  \BibitemOpen
  \bibfield  {author} {\bibinfo {author} {\bibfnamefont {N.~J.}\ \bibnamefont {Cornish}},\ }\href@noop {} {\bibfield  {journal} {\bibinfo  {journal} {Physical Review D}\ }\textbf {\bibinfo {volume} {105}},\ \bibinfo {pages} {044007} (\bibinfo {year} {2022})}\BibitemShut {NoStop}%
\end{thebibliography}%

\end{document}